\documentclass[aps,preprint]{revtex4}

\usepackage{graphicx}

\newcommand\beq{\begin{equation}}
\newcommand\eeq{\end{equation}}
\newcommand\beqa{\begin{eqnarray}}
\newcommand\eeqa{\end{eqnarray}}

\newcommand{\ds}[1]{#1 \hspace{-0.5em}/}  

\newcommand\bgamma{\mbox{\boldmath$\gamma$}}

\begin{document}
\title{
Dual chiral density wave in quark matter}
\author{T.Tatsumi$^a$ and E.Nakano$^b$}

\affiliation{${^a}${\it Department of Physics, Kyoto University, 
                Kyoto 606-8502, Japan} \\
         ${^b}${\it Department of Physics, Tokyo Metropolitan University, 
         1-1 Minami-Ohsawa, Hachioji, Tokyo 192-0397, Japan }}

\date{\today}

\begin{abstract}
Possible manifestation of a {\it dual chiral density wave} in quark
 matter is discussed in relation to the chiral symmetry of QCD, 
which is described by a dual standing wave of 
the scalar and pseudo-scalar densities.
It is demonstrated  that quark matter is unstable for forming 
the dual chiral density wave at moderate densities 
in the Nambu-Jona-Lasinio model; accordingly the critical density for restoration 
of chiral symmetry becomes higher than ever considered.
A unique magnetic property is also pointed out in the new phase.
\vskip 0.5cm
\noindent PACS: 11.30.Rd; 75.10.-b; 26.60.+c\\
Keywords: quark matter; magnetism; chiral symmetry; 
dual chiral density wave 
\vfill
\noindent -------------------------

\noindent E-mail addresses: tatsumi@ruby.scphys.kyoto-u.ac.jp (T. Tatsumi); 
enakano@comp.metro-u.ac.jp (E. Nakano)
\end{abstract}

\maketitle

Recently condensed matter physics of QCD has been an exciting area in 
nuclear physics. 
Superconductor model of the vacuum is a classic one, firstly 
considered by Nambu and Jona-Lasinio to understand the nucleon mass 
\cite{nam}. Nowadays it is believed that quark-anti-quark pairs
condense in the vacuum, leading to the spontaneous symmetry breaking
(SSB) of chiral
symmetry in QCD.  In the past decade 
the color superconductivity (quark-quark pair condensation) 
in quark matter has been 
extensively studied by many authors \cite{bai}. 
There are also some works to inquire the 
coexistence of these two phases on the phase diagram of QCD  
\cite{ber}. 

On the other hand, the possibility of 
ferromagnetism (FM) in quark matter has been also discussed
\cite{tat,ferr1}, stimulated by the discovery of magnetars \cite{woo}. 
It has been suggested \cite{tat} that ferromagnetic phase exists at 
rather low densities in analogy with an electron gas discussed by Bloch 
\cite{blo} and demonstrated that relativistic effects play an important role 
in quark matter. We have also studied a coexistent mechanism of FM and
CSC by assuming the spin-parallel pairing within the mean-filed
approximation \cite{nak}.

It should be noted that 
CSC arises from particle-particle ({\it p-p}) correlation, while
particle-hole ({\it p-h}) correlations should be responsible for
SSB of chiral symmetry and FM. 
Besides, different types of the {\it p-h} condensations has been 
 proposed \cite{der1, der2} in relation to chiral symmetry, 
where the scalar or tensor density  forms 
a standing wave with a finite momentum, 
called the chiral density wave. 
The instability for the density wave in quark matter was first discussed 
by Deryagin {\it et al.} \cite{der0} at asymptotically high densities 
where the interaction is very weak, 
and they concluded that the density-wave instability prevails over the BCS one 
in the large $N_c$ (the number of colors) limit 
due to the dynamical suppression of colored pairing interactions. 
 
We consider here another type of the density wave
due to chiral symmetry in QCD. 
In this letter we are concentrated 
on the flavor-$SU(2)$ case, and 
study the flavor-singlet scalar density, 
$\langle\bar q q\rangle$, 
and the flavor-triplet pseudo-scalar density, 
$\langle\bar q i\gamma_5\tau_iq\rangle$, 
which transform as the $[1/2,1/2]$ representation 
in $SU(2)_L\times SU(2)_R$ chiral symmetry.
In the SSB phase, 
they satisfy the constraint, $\langle\bar q
q\rangle^2+\langle\bar q i\gamma_5\tau_iq\rangle^2=A^2$ in the chiral limit 
within the mean-field approximation, and  
the ground states are degenerate on the hypersphere in the chiral space, 
prescribed by the scalar and pseudo-scalar mean-fields, 
if both densities are spatial constants. 

We introduce  
a {\it dual chiral-density wave} (DCDW) state, where both the scalar and
pseudo-scalar densities always 
reside on the hypersphere with a constant modulus $A$, 
while each density is {\it spatially non-uniform}; 
restricting ourselves to the electric charge eigenstate, 
we consider the following configuration in quark matter,
\begin{eqnarray}
\langle\bar q q\rangle &=& A\cos\theta({\bf r})\nonumber\\
\langle\bar q i\gamma_5\tau_3q\rangle&=&A\sin\theta({\bf r}).
\label{gene1}
\end{eqnarray}
The chiral angle $\theta$ represents the degree of freedom 
of the Nambu-Goldstone mode or the directional mode on the hypersphere.  
Taking the simplest but  nontrivial form for the 
chiral angle $\theta$ such that $\theta({\bf r})={\bf q\cdot r}$,
we call this configuration DCDW.  

It should be noted that 
we can construct the DCDW state by acting a space-dependent 
chiral transformation such as 
$
q \longrightarrow \exp(i\gamma_5\theta({\bf r})/2)q,
$
on the usual SSB phase 
where only the scalar density condenses. 
This is nothing but a kind of Weinberg transformation \cite{wei}.

When the chiral angle has some spatial dependence, 
there should appear two extra terms in the effective potential
as consequences of the local chiral transformation: 
one is the interaction term, 
$\bar{q} (\gamma_5\bgamma\cdot\nabla\theta/2) q$, 
due to the non-commutability of $\theta({\bf r})$ with the kinetic
(differential) operator in the Dirac operator. 
 It may be easily seen
that $\nabla\theta$ acts as an ``external'' axial-vector field for
quarks and resolves the degeneracy of the single-particle energy spectra for
different spin states. In this case, by a suitable rearrangement of the two Fermi seas
we can always find an optimal configuration to lower the total energy.
Another one is nontrivial and comes from the 
vacuum-polarization effect: the energy spectrum of quarks is modified in the
presence of $\theta({\bf r})$ and thereby the vacuum polarization should
give an 
additional term, $\propto (\nabla\theta)^2$ in the lowest order. 
This can be regarded as an appearance of the kinetic energy term for
DCDW through the vacuum polarization \cite{sug}.  
Thus, the Fermi sea works for DCDW, while the Dirac sea against it; 
when the interaction energy is superior to the kinetic energy, quark
matter becomes unstable to form DCDW.

We will see that 
the mechanism is quite similar to that for the spin density wave 
suggested by Overhauser \cite{ove,sdw1}, and entirely reflects many-body effects. In general, 
such a  density wave is favored in 1-D (one spatial dimension) systems 
with the wave number $|{\bf q}|=2p_F$ ($p_F:$ Fermi momentum) according to the Peierls instability 
\cite{peiel1}, 
e.g., charge density waves in quasi-1-D metals \cite{peiel2}. 
The essence of its mechanism is the nesting of Fermi surfaces 
and the 
level crossing of the  single-particle energy spectra with a relative momentum 
 ${\bf q}$. 
In the higher dimensional systems, however, the nesting is incomplete
and the density wave should be
formed 
provided the interaction of a relevant ({\it p-h}) channel is strong enough. 
For the 3-D electron gas, 
it was shown by Overhauser \cite{ove,sdw1} that the paramagnetic state is unstable 
with respect to the formation of the static spin-density wave, 
in which the energy spectra of up- and down-spin states with a finite relative
momentum exhibit a level crossing and it is resolved by 
the spin exchange interaction, 
while the wave number does not coincide with $2p_F$ completely 
because of the higher dimension.

We explicitly demonstrate that quark matter 
becomes unstable for a formation of DCDW at moderate densities, 
using the Nambu-Jona-Lasinio (NJL) model 
which is originally one of the schematic models 
to describe realization of chiral symmetry in the vacuum \cite{nam}. 
Recently the model has been also considered as 
an effective model of QCD embodying SSB of chiral symmetry 
in terms of quark degree of freedom \cite{kle}. 

We start with the NJL Lagrangian with $N_f=2$ flavors and $N_c=3$ colors,
\beq
{\cal L}_{NJL}
=\bar\psi(i\ds{\partial}-m_c)\psi+G[(\bar\psi\psi)^2+
(\bar\psi i\gamma_5\mbox{\boldmath$\tau$}\psi)^2],
\label{njl}
\eeq
where $m_c$ is the current mass, $m_c\simeq 5$MeV.
Under the mean-field approximation, 
we keep the pseudo-scalar as well as the scalar mean-field. 
In the usual treatment to generate a dynamical quark mass and 
see the restoration of chiral symmetry within the NJL model,  
the pseudo-scalar mean-field is implicitly discarded. 
However, note that this is justified only for the vacuum 
since it is the definite eigenstate of parity, 
and there is no compelling reason at finite baryon density. 
We assume here the following form for the mean-fields,
\beqa
\langle\bar\psi\psi\rangle&=&\Delta\cos({\bf q}\cdot {\bf r}) \nonumber\\
\langle\bar\psi i\gamma_5\tau_3\psi\rangle&=&
\Delta\sin({\bf q}\cdot {\bf r}), 
\label{chiral}
\eeqa
in the direct channel. Accordingly, 
we define a new quark field $\psi_W$ by the Weinberg 
transformation,
\beq
\psi_W=\exp[i\gamma_5\tau_3 {\bf q} \cdot {\bf r}/2 ]\psi,
\label{wein}
\eeq
to separate the degrees of freedom of the amplitude and the phase of  
DCDW in the Lagrangian. 
In terms of the new field the effective Lagrangian 
renders
\beq
{\cal L}_{MF}=\bar\psi_W[i\ds{\partial}-M-1/2\gamma_5\tau_3\ds{q}]\psi_W
-G\Delta^2+{\cal L}_{\rm SB},
\label{effl}
\eeq
where ${\cal L}_{\rm SB}$ is a small residual term due to the current 
mass $m_c$, \\
${\cal L}_{\rm SB}
=-m_c\bar\psi_W[\exp(i\gamma_5\tau_3 {\bf q}\cdot {\bf r})-1]\psi_W$, 
and we put $M\equiv m_c-2G\Delta$ and $q^\mu=(0, {\bf q})$. In the 
following, we 
take the chiral limit ($m_c=0$) discarding ${\cal L}_{\rm SB}$.  
The form given in (\ref{effl}) appears to be the same as 
that in the superconductor model of the vacuum except an ``external'' 
axial-vector field, ${\bf q}$,  
generated by the wave vector of DCDW \cite{kle}; the {\it amplitude} of DCDW 
generates the dynamical quark mass in this case. Accordingly we can see 
that our theory becomes trivial in the chiral-symmetry restored phase.

The Dirac equation for $\psi_W$ then gives a 
{\it spatially uniform} solution,
$\psi_W=u_W(p)\exp(i{\bf p\cdot r})$, 
with the eigenvalues 
\beq
E^{\pm}({\bf p})=\sqrt{E_{p}^{2}+|{\bf q}|^2/4\pm \sqrt{({\bf
p}\cdot{\bf q})^2+M^{2}|{\bf q}|^2}},~~~E_p=(M^2+|{\bf p}|^2)^{1/2}
\label{energy}
\eeq
for positive energy (valence) quarks with different polarizations
denoted by the sign $\pm$. For 
negative energy quarks in the Dirac sea, they have an energy spectrum symmetric with 
respect to null line because of charge conjugation symmetry in the  
Lagrangian (\ref{effl}).

The single-particle energy spectra (\ref{energy}) exhibit an analogue of the 
{\it exchange splitting}
between energies with different spin states in the Stoner model
\cite{blo}. 
Hereafter, we choose ${\bf q}//\hat z$, ${\bf q}=(0,0,q)$ with $q>0$, 
without loss of generality. Note that we need not distinguish two flavors 
in the energy spectra, 
while their eigenspinors are different to each other.
The energy spectra (\ref{energy}) show a salient feature: 
they break  
rotation symmetry due to the couping term of the momentum and the wave
vector, and lead to the axially-symmetric deformed Fermi seas.

The effective potential then reads
\beqa
\Omega_{\rm total}&=&\gamma\int\frac{d^3p}{(2\pi)^3}
\left[(E^-_p-\mu)\theta_-+(E^+_p-\mu)\theta_+\right]
-\gamma\int\frac{d^3p}{(2\pi)^3}\left[E^-_p+E^+_p\right]
+(M-m_c)^2/4G\nonumber\\
&\equiv&\Omega_{\rm val}+\Omega_{\rm vac}+M^2/4G 
\label{therm}
\eeqa
after summing up all the energy levels, where $\theta_\pm=\theta(\mu-E^\pm_p)$, $\mu$ is the chemical potential and 
$\gamma$ the degeneracy factor $\gamma=N_fN_c(=6)$. The first term 
$\Omega_{\rm val}$ is the 
contribution by the valence quarks filled up to the chemical potential
$\mu$ in each Fermi sea, 
while the second term $\Omega_{\rm vac}$ is the vacuum 
contribution that is formally divergent. We shall see both contributions
are {\it indispensable} in our context.  
Since the NJL model is nonrenormalizable, 
we need some regularization procedure 
to extract a meaningful finite value for the vacuum contribution.
Note that we cannot apply the usual momentum cut-off 
regularization (MCOR) scheme to $\Omega_{\rm{vac}}$, 
since the energy spectrum has no more rotation symmetry. 
Instead, we adopt the proper-time regularization (PTR) scheme \cite{sch}, 
which may be a most suitable one for our purpose, 
since $\Omega_{\rm{vac}}$ counts the vacuum-polarization contributions under 
the ``external'' axial-vector field ${\bf q}$.
Incidentally the vacuum polarization effect   
under the external electromagnetic field has been treated in a gauge 
invariant way by using the PTR scheme, where the energy spectrum is also deformed depending on the field strength \cite{sch}.

The vacuum contribution $\Omega_{\rm vac}$ can be represented as 
\beq
\Omega_{\rm vac} = i\gamma\int\frac{d^4p}{(2\pi)^4}
{\rm tr}{\rm ln}(S_W(p)/S_0(p)), 
\label{e}
\eeq
where $S_W(p)$ is the quark propagator,
$S_W(p)=[\ds{p}-M-1/2\gamma_5\tau_3\ds{q}]^{-1}$. 
Here we subtracted an irrelevant constant 
$\Omega_{\rm ref}=i\gamma\int d^4p/(2\pi)^4 {\rm tr}{\rm ln}(S_0(p))$ 
with $S_0(p)=[\ds{p}-m_{\rm ref}]^{-1}$ to make the following 
procedure mathematically well-defined, which is the quark propagator with
an arbitrary mass $m_{\rm ref}$ .
Introducing the proper-time variable $\tau$, we eventually find
\beq
\Omega_{\rm vac}=\frac{\gamma}{8\pi^{3/2}}\int_0^\infty
\frac{d\tau}{\tau^{5/2}}
\int^\infty_{-\infty}\frac{dp_z}{2\pi}\left[
e^{-(\sqrt{p_z^2+M^2}+q/2)^2\tau}
+e^{-(\sqrt{p_z^2+M^2}-q/2)^2\tau}\right]-\Omega_{\rm ref},
\label{j}
\eeq
which is obviously reduced to the standard formula 
\cite{kle} in the limit $q\rightarrow 0$ (no DCDW). 
The integral with respect to $\tau$ is not well defined yet as it is,     
since it is still divergent 
due to the $\tau\simeq 0$ contribution.
Regularization further proceeds by replacing the lower bound of the integration range 
by $1/\Lambda^2$, which corresponds to the momentum cut-off in the 
MCOR scheme.

Now we examine a possible instability of quark matter with respect to formation 
of DCDW. For the vacuum contribution, we can easily see 
$ \Omega_{\rm vac}\geq \Omega_{\rm vac}|_{q \rightarrow 0}$, 
which means the vacuum (the Dirac sea) is stable against formation
of DCDW, as it should be. 
For small $q$, 
\beqa
\Omega_{\rm vac}
&=& \Omega_{\rm vac}|_{q \rightarrow 0} 
+\frac{\gamma\Lambda^2}{16\pi^2}J(M^2/\Lambda^2) q^2+O(q^4) 
\label{l}
\eeqa
where $J(x)$ is a function,
$
J(x)=-x{\rm Ei}(-x),
$
with the exponential integral ${\rm Ei}(-x)$.
The coefficient of $q^2$ term in 
$\Omega_{\rm vac}$, which is called the spin stiffness 
parameter $\beta_{\rm vac}$, 
has a definite 
physical meaning: since the pion decay 
constant $f_\pi$ is given in terms of $J(x)$ 
within the NJL model \cite{kle},
\beq
f_\pi^2=\frac{\gamma\Lambda^2}{8\pi^2}J(M^2/\Lambda^2),
\label{fpi}
\eeq 
$\beta_{\rm vac}$ can be written as
$
\beta_{\rm vac}=\frac{1}{2}f_\pi^2.
$ 
There are two remarks in order. One is that there appears no divergence as 
$\Lambda\rightarrow\infty$ in the higher order terms in $q^2$. 
Hence, 
strictly speaking, higher order terms should be discarded \cite{sug}, since 
the divergent term should dominate over other finite terms. However, we  
keep them here to preserve self-consistency after introduction of the 
regularization in the nonrenormalizable theory.  
The other 
is that Eq.~(\ref{l}) suggests that the vacuum-polarization effect precisely 
provides the expected kinetic-energy term for DCDW, except 
a ``running'' pion decay constant (\ref{fpi}).

For given $\mu, M$ and $q$ we can evaluate the valence contribution 
$\Omega_{\rm val}$ using 
Eq.~(\ref{energy}) and write down the general formula 
analytically. For small $q$ we have a following expansion,
\beqa
\Omega_{\rm val}=\Omega_{\rm val}|_{q\rightarrow 0}
-\frac{\gamma}{8\pi^2}M^2q^2H(\mu/M)+O(q^4)
\label{xb}
\eeqa 
where  $\Omega_{\rm val}|_{q\rightarrow 0}
=\gamma/(24\pi^2)\left[\mu\sqrt{\mu^2-M^2}
\left(5M^2-2\mu^2\right)
-3M^4{\rm ln}(\mu+\sqrt{\mu^2-M^2}/m)\right]$ for normal
quark matter and $H(x)={\rm ln}(x+\sqrt{x^2-1})$.
The spin stiffness parameter then reads
\beq
\beta_{\rm val}=-\frac{\gamma}{8\pi^2}M^2H(\mu/M)
\eeq
for valence quarks. Since the function $H(x)$ is always positive and 
a monotonously increasing function, we can see $\beta_{\rm val}\leq 0$. 

The total spin-stiffness parameter, 
$\beta_{\rm tot} \equiv \beta_{\rm val}+\beta_{\rm vac}$, 
depends on the dynamical mass and the chemical potential for a given $\Lambda$.
In the vacuum it is positive, and the necessary condition to form DCDW
then reads $\beta_{\rm tot}<0$ due to the finite-density effect. 
When this condition is fulfilled, 
the optimal value of $q$ is determined by the minimum of 
the effective potential (\ref{therm}).  
We can see that the optimal value of $q$ is $O(2p_F)$ with the quark Fermi
momentum $p_F$ without recourse to explicit calculations. 
First, consider the energy spectra for massless quarks (see Fig.1). 
\begin{figure}[h]
\begin{center}
\includegraphics[width=0.4\textwidth,keepaspectratio]{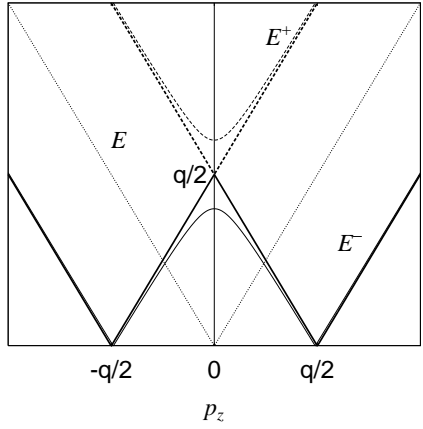}
\end{center}
 \caption{Energy spectra for ${\bf p}_{\perp}=0$. $E^\pm$ with $M=0$ 
(thick solid and dashed lines).
$\tilde E^\pm$ with the 
definite chirality is also shown for comparison (dotted line). 
We can see there is a degeneracy of $E^\pm$ 
at $p_z=0$ for $M=0$, while it is resolved by the mass 
(thin solid and dashed lines).}
\end{figure}
As is already discussed, 
our theory becomes trivial in this case 
and we find two energy spectra   
\beq
E^\pm({\bf p})=\sqrt{p_\perp^2+(|p_z|\pm q/2)^2}, ~~~{\bf p}_\perp=(p_x,p_y, 0),
\label{zeom}
\eeq
essentially equivalent to the usual ones $E^\pm({\bf p})=|{\bf p}|(\equiv
E({\bf p}))$; taking a linear 
combination of the eigenspinors $\psi_W^\pm$, we can reconstruct the 
eigenstates of definite chirality $\gamma_5$ to get the energy spectra, 
$\tilde E^\pm({\bf p})=\sqrt{p_\perp^2+(p_z\pm q/2)^2}$. Then we can see
a level crossing on the  $p_z=0$ plane. Once the mass term is taken into 
account this degeneracy is resolved and the energy gap appears there. 
The resultant spectra are given by (\ref{energy}). Hence  we have always an energy gain by filling only the
lower-energy spectrum $E^-({\bf p})$ up to the Fermi energy, if the relation 
,$q=O(2p_F)$, holds.
Thus, we find that this  
mechanism is very similar to that of spin density wave 
by Overhauser \cite{ove,sdw1}
\footnote{It has been shown that similar mechanism by valence quarks
also works in another context of the chiral density 
wave \cite{der1, der2}.}
.

\begin{figure}[h]
\begin{center}
\includegraphics[height=6cm]{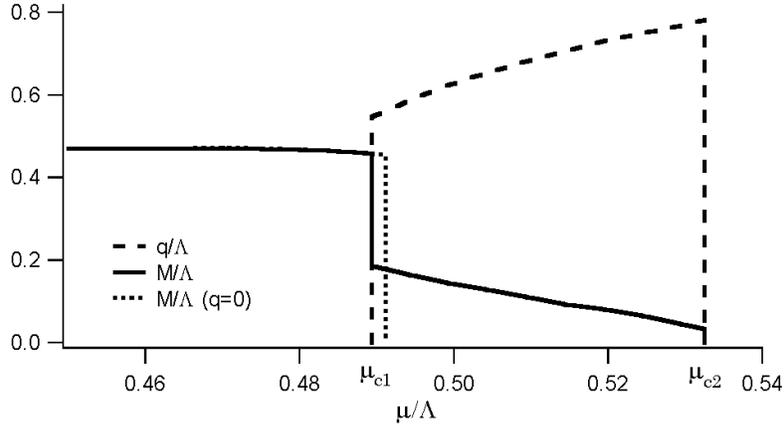}
\end{center}
\caption{The wave number $q$ and the dynamical mass $M$ are plotted 
as functions of the chemical potential at zero temperature. 
Solid (dotted) line for $M$ with (without) DCDW, 
and dashed line for $q$.}
\label{op1}
\end{figure}

Taking the extremum of the effective potential (\ref{therm}) with respect to the
order-parameters $M$ and $q$, we can determine their values for given
baryon-number densities. 
Fig.~\ref{op1} demonstrates the behaviors of the optimum values  
as functions of $\mu$ at zero temperature 
for a parameter set, $G \Lambda^2=6, \Lambda=860$ MeV.
%
%
It is found that the magnitude of $q$ becomes finite 
just before the ordinary chiral-symmetry restoration, 
and DCDW survives in the finite range of $\mu$, 
$\mu_{c1}\le\mu\le\mu_{c2}$,  
which corresponds to the baryon-number densities $\rho_b/\rho_0=3.62-5.30$. 
The dynamical mass remains finite in this density region, which means
chiral restoration is delayed in the presence of DCDW.  
The wave number $q$ increases with $\mu$, 
while its value is smaller than
$2p_F$($\simeq 2\mu$ for massless quarks) 
due to the higher dimensional effects. 
Actually, the ratio of the wave number and the Fermi momentum 
(at normal phase $q=M=0$) lies in the range, $q/p_F=1.17-1.47$.   
We thus conclude that DCDW is induced by finite-density contributions, 
and has the effect to extend the SSB phase ($M\neq 0$) to high density
region, which suggests another path for chiral-symmetry restoration by
way of the DCDW state at finite density.

Here we would like to point out an interesting aspect of the DCDW state. 
The quark-DCDW coupling is spin dependent and 
we can see that it gives rise to a spatial oscillation of the magnetic moment
in quark matter. The magnetic moment is given as
\beqa
M_z &\propto& \langle\bar \psi\sigma_{12}\psi\rangle 
=M_z^W\cos({\bf q}\cdot {\bf r}),
\eeqa
where $M_z^W (=\langle\bar \psi_W\sigma_{12}\psi_W\rangle) $ is 
proportional to the dynamical mass. 
Thus DCDW can be regarded as a kind of the 
{\it spin density wave} \cite{ove,sdw1} 
and should have phenomenological implications on 
the magnetic properties of compact stars.

In this Letter we have only considered the direct channel (Hartree term) 
of the four Fermi 
interaction. When we consider the exchange contributions 
(Fock term), there appear additional quark-quark interactions 
in the vector and 
axial-vector channels by way of the Fierz transformation
\cite{nak, mar}. 
It is well known that the vector contribution 
is rather trivial and summed up as a renormalization of the chemical 
potential. We 
are then interested in the axial-vector contribution in our context. 
We may see 
that it renormalizes the wave vector of DCDW; 
this can be regarded as a kind of vertex renormalization 
of the quark-DCDW coupling. 
This subject will be discussed elsewhere \cite{naktat}. 

It would be interesting to recall that DCDW  
is similar to pion condensation within the $\sigma$ model, 
considered by Dautry and Nyman \cite{dau}, where $\sigma$ and $\pi^0$ 
meson condensates take the same form as Eq.~(\ref{gene1}). So it might
be possible to connect pion condensation before deconfinement with DCDW
after it by a symmetry consideration. However, the restoration of chiral
symmetry, especially the vacuum contribution, has been poorly considered 
in the context of pion condensation.

Finally, it might be interesting to generalize the DCDW configuration
(\ref{chiral}) by
incorporating the radial degree of freedom; one may generalize the amplitude to be
spatially dependent as well \'a la refs.\cite{der1,der2}, but one must consider a
genuine {\it non-uniform} matter in this case. Anyhow, if two degrees of
freedom work coherently, we may have more favorable configurations than
considered here.

We thank T. Maruyama, H. Yabu and K. Nawa for discussions. 
This work is supported by the Grant-in-Aid for the 21st Century COE
``Center for the Diversity and Universality in Physics '' from 
the Ministry of Education, Culture, Sports, Science and
Technology of Japan. It is also supported by the Japanese 
Grant-in-Aid for Scientific
Research Fund of the Ministry of Education, Culture, Sports, Science and
Technology (13640282, 16540246).


\begin{thebibliography}{99}
\bibitem{nam}
Y. Nambu and G. Jona-Lasinio, Phys. Rev. {\bf 122} (1961) 345. 

\bibitem{bai}
D. Bailin and A. Love, Phys. Reports 107(1984) 325.\\
For a review, e.g., M. Alford, 
Ann. Rev. Nucl. Part. Sci. 51 (2001) 131.

\bibitem{ber} 
J. Berges and K. Rajagopal, Nucl. Phys. {\bf B538} (1999) 215. \\
R. Rapp, T. Schaefer, E.V. Shuryak, M. Velkovsky, 
Phys. Rev. Lett. 81 (1998) 53. 

\bibitem{tat}
T. Tatsumi, Phys. Lett. {\bf B489} (2000) 280.

\bibitem{ferr1} A. Ni\'{e}gawa, hep-ph/0404252. 

\bibitem{woo} For a recent review, P.M. Woods and C. Thompson, astro-ph/0406133.

\bibitem{blo} 
F. Bloch, Z. Phys. {\bf 57} (1929) 545.\\
K. Yosida, {\it Theory of magnetism} (Springer, 1998).

\bibitem{nak} E. Nakano, T. Maruyama and T. Tatsumi, 
Phys. Rev. {\bf D68} (2003) 105001.

\bibitem{der1}
E. Shuster and D. T. Son, Nucl. Phys. {\bf B}573 (2000) 434. 

\bibitem{der2}
B.-Y. Park, M.Rho, A.Wirzba and I.Zahed, Phys. Rev. {\bf D62} (2000) 034015.\\
R. Rapp, E.Shuryak and I. Zahed, Phys. Rev. {\bf D63} (2001) 034008. 

\bibitem{der0}
D.V. Deryagin, D. Yu. Grigoriev and V.A. Rubakov, 
Int. J. Mod. Phys. {\bf A7} (1992) 659. 

\bibitem{wei} 
S. Weinberg, {\it The quantum theory of field II} (Cambridge, 1996).

\bibitem{sug}
T. Eguchi and H. Sugawara, Phys. Rev. {\bf D10} (1974) 4257.\\
K. Kikkawa, Prog. Theor. Phys. {\bf 56} (1974) 947.


\bibitem{peiel1}
R. E. Peierls, {\it Quantum Theory of Solids} 
(Oxford University Press, London, 1955). 

\bibitem{peiel2}
G. Gr\"{u}ner, Rev. Mod. Phys. {\bf 60} (1988) 4. 

\bibitem{ove}
A.W. Overhauser,  Phys. Rev. Lett.{\bf 4} (1960) 462;   
                  Phys. Rev. {\bf 128} (1962) 1437. 

\bibitem{sdw1}
G. Gr\"{u}ner, Rev. Mod. Phys. {\bf 66} (1994) 1. 

\bibitem{kle}
S.P. Klevansky, Rev. Mod. Phys. {\bf 64} (1992) 649.\\
T. Hatsuda and T. Kunihiro, Phys. Rep. {\bf 247} (1994) 221.

\bibitem{sch}
J. Schwinger, Phys. Rev. {\bf 92} (1951) 664.

\bibitem{mar}
T. Maruyama and T. Tatsumi, Nucl. Phys. {\bf A693} (2001) 710.

\bibitem{naktat} E. Nakano and T. Tatsumi, in preparation. 

\bibitem{dau}
F. Dautry and E.M. Nyman, Nucl. Phys. {\bf 319} (1979) 323.\\
K. Takahashi and T. Tatsumi, Phys. Rev. {\bf C63} (2000) 015205;
Prog. Theor. Phys. {\bf 105} (2001) 437. \\ 
M. Kutschera, W. Broniowski and A. Kotlorz, Nucl. Phys. 
{\bf A516} (1990) 566.

\end{thebibliography}
\end{document}